\documentclass[cits]{PoS}

\usepackage{array}
\newcolumntype{x}[1]{%
>{\centering\hspace{0pt}}p{#1}}%
\newcommand{\Msun}{M$_{\odot}\,$}

\title{Unveiling Physical Processes in Type Ia Supernovae with a Laue Lens Telescope}

\ShortTitle{Unveiling Physical Processes in Type Ia Supernovae with a Laue Lens Telescope}

\author{\speaker{N. M. Barri\`{e}re}, J. A. Tomsick, S. E. Boggs\\
        Space Sciences Laboratory, University of California Berkeley, 7 Gauss Way, Berkeley CA 94720-7450, USA \\
        E-mail: \email{barriere@ssl.berkeley.edu}
        }

\author{P. von Ballmoos, J. Rousselle\\
        Centre d'Etude Spatiale des Rayonnements, UMR 5187, 9, av du Colonel Roche, 31028 Toulouse, France\\
        }

\abstract{We present in this paper a focusing gamma-ray telescope with the goal of addressing the true nature of Type Ia Supernovae (SNe Ia). This telescope is based on a Laue lens focusing a 100 keV wide energy band centered on 847 keV, which corresponds to a bright line emitted by the decay chain of $^{56}$Ni, a radioactive element massively produced during SNe Ia events.  Spectroscopy and light curve measurements of this gamma-ray line allow for a direct measurement of the underlying explosion physics and dynamics, and thus discriminate among the competing models. However, reaching this goal requires the observation of several events with high detection significance, meaning more powerful telescopes.  The telescope concept we present here is composed of a Laue lens held 30 m from the focal plane instrument (a compact Compton telescope) by an extendible mast. With a 3-$\sigma$ sensitivity of 1.8$\times$10$^{-6}$ ph/s/cm$^2$ in the 3\%-broadened line at 847 keV (in 1 Ms observation time), dozens of SNe Ia could be detected per year out to $\sim$40 Mpc, enough to perform detailed time-evolved spectroscopy on several events each year.  This study took place in the framework of the DUAL mission proposal which was recently submitted to ESA for the third medium class mission of the Cosmic Vision program.}

\FullConference{8th INTEGRAL Workshop ``The Restless Gamma-ray Universe''- Integral2010,\\
		September 27-30, 2010\\
		Dublin Ireland}

\begin{document}

\section{Introduction}

Type Ia supernovae (SNe Ia) are used as a cosmological standard candle to determine extragalactic distances with ever increasing precision. Based on this, extensive efforts during the past two decades led to the astonishing result that the expansion of the Universe appears to be accelerating, which implies the existence of a dark energy \cite{riess.1998fk, perlmutter.1999uq}. Despite the fact that this result is apparently sound, it is well known that SNe Ia are actually not standard candles. Their intrinsic luminosity can vary over a factor of three, as was emphasized for instance in 1991 with the extremely faint SN 1991bg \cite{filippenko.1992vn} and the extremely bright SN 1991T \cite{phillips.1992ys}. However, Phillips (1993)\cite{phillips.1993kx} showed that there is a correlation between the shape of a SN Ia light curve and its luminosity: Supernovae with the steepest decline are the least luminous. This one-parameter empirical law proved to work extremely well, but does not provide any explanation for why it is the case. Why SNe Ia brightnesses change with light curve shape is a mystery, which is actually related to our lack of understanding of the supernova explosions themselves. Without even knowing the progenitor system of these objects, it is difficult to infer any observational feature. Whereas a deeper understanding of SNe Ia physics is not likely to threaten the outstanding cosmological results obtained during the last two decades, it would help guide the empirical work, and provide confidence that stellar evolution is not subtly undermining the cosmological inferences. More importantly, it would help reduce the systematics associated with their use as distance indicator. Aside from cosmology, SNe Ia are actually a full-fledged topic of study on their own, as they are laboratories for extreme physics. Understanding their underlying physical processes would also clarify their role in the nucleosynthesis of the heavy elements present in the Universe

Future ground-based observatories and space-borne missions promise to bring thousands of new supernova discoveries that will allow their classification into subcategories with significant statistics. Sorting SNe Ia by, for instance, host galaxy type, host galaxy metallically, color and redshift, will certainly uncover new correlations and allow us to infer some of their intrinsic properties. Observations in the gamma-ray domain cannot bring such huge statistics. Even the most sensitive instruments such as those presented in this paper can ``only'' detect a few tens of SNe Ia per year. However, the observation of the gamma-ray lines emitted by the decay of radioactive elements synthesized during the supernova explosion is a very powerful probe of the physics of these objets. All SN Ia models predict that a large quantity of radioactive $^{56}$Ni is synthesized, which decays following the $^{56}$Ni $\rightarrow$ $^{56}$Co $\rightarrow$ $^{56}$Fe chain, emitting many gamma-ray lines including the bright line at 847 keV. The observation of this line reveals the location of the radioactivity within the ejecta through the time-dependence of the photon escape, and the ejection velocities of various layers through the line Doppler profiles. Therefore, spectroscopy and light curve measurements of the gamma-ray lines allow for direct measurement of the underlying explosion physics and dynamics, and thus discriminate among the competing models \cite{pinto.2001kx}.

There are currently three classes of competing models to describe SNe Ia, all of which feature a degenerate C/O white dwarf (WD) in a binary system. A C/O WD is the final state of a main sequence star whose mass was in the 0.5 - 6.5 \Msun  range, which allows the fusion of He into C and O via the triple-alpha process, but not the fusion of C into Ne, thus producing an inert core of C and O. After the star has ejected its remaining He and H layers into a planetary nebula, the C/O WD remains and has a maximum mass of $\sim$ 1.1 - 1.2 \Msun \cite{weidemann.2000kx}.

The most commonly accepted scenario to produce a SN Ia involves a massive C/O WD accreting matter from a companion star in close binary system. As the mass of the WD builds up and reaches the Chandraskhar mass limit ($\sim$ 1.38 \Msun), its central density rises above 10$^9$ g cm$^{-3}$ resulting in the ignition of C fusion near its center. The outward-propagating thermonuclear flame quickly becomes turbulent, causing an energetic explosion. Although this is the ``standard model'', this scenario is challenged by the fact that copious X-ray emission is expected during an accretion phase that should last about 10$^7$ years, but it is not observed. This fact suggests that this scenario might not be the dominant one \cite{gilfanov.2010uq}.

An alternative or perhaps a complementary class of models involves a C/O WD of mass in the range 0.6-0.9 \Msun bound to a main sequence star. Like in the first scenario, the companion star feeds the WD with its matter, mostly H and He. After a certain time, fusion reactions ignite at the base of the accreted layer producing an inward propagating compression wave, driving the central density and temperature in a range that allows for carbon fusion under explosive conditions. In this sub-Chandrasekhar scenario, the amount of $^{56}$Ni produced increases with the mass of the WD. This could provide an explanation for the brightness decline rate relationship described by Phillips (1993), as more $^{56}$Ni implies brighter SNe Ia and a longer diffusion time \cite{pinto.2001kx}.

The third class of models involves the coalescence of two C/O WD where the orbital distance decays by emission of gravitational waves \cite{mochkovitch.1990vn}. While there are several possibilities for bringing the two WD together (for instance, through an accretion disc formed by the less massive WD or through a direct collision), the result is a massive C/O WD with a mass exceeding the Chandrasekhar limit in most cases. At this point, the story is the same as in the first scenario.  This model is supported by recent observational hints; For instance, Scalzo et al. (2010)\cite{scalzo.2010zr} reports that SN 2007if produced $\sim$1.6 \Msun of $^{56}$Ni, and Tanaka et al. (2010) \cite{tanaka.2010ys} suggest that SN 2009dc produced $\gtrsim$ 1.2 \Msun of $^{56}$Ni. In both cases, the progenitor should have had a mass exceeding the Chandrasekhar limit.

In order to discriminate between these models, detections with high significance ($\gtrsim 25 \sigma$) are needed to have enough counts to obtain detailed spectra. The requirement is to detect \emph{at least} one SN Ia per year with such photon statistics. However, past and present instruments were not sensitive enough to reach this goal, mainly due to the fact that they were not using focusing optics. Both coded mask and Compton telescope share the same problem, their instrumental background is (roughly) proportional to their volume, which means that their sensitivity scales only with the square root of their exposed area. Focusing instruments have two tremendous advantages: first, the volume of the focal plane detector can be made much smaller than for non-focusing instruments, and second, the residual background, often time-variable, can be measured simultaneously with the source, and can be reliably subtracted. The concept of a Laue gamma-ray lens holds the promise of extending focusing capabilities into the MeV range. In order to achieve the ultimate sensitivity for the gamma-ray lens mission, the focal plane detector must be designed to match the characteristics of the lens focal spot. A Compton telescope is a good solution because the focal plane is intrinsically finely pixelated, optimized for MeV gamma-ray detection, and because the direction of incident gamma-rays can be determined by the Compton reconstruction to enable discrimination gamma-rays coming from the lens (``electronic collimation'').

In this paper, we describe a Laue lens gamma-ray telescope concept dedicated to the study of SNe Ia through the observation of the 847 keV line. This work has been done in the framework of the DUAL mission \cite{von-ballmoos.2010zr, boggs.2010uq}, which has recently been proposed to the European Space Agency as a medium class mission. In the next section, we introduce the basics of Laue lenses. Then, the telescope concept is presented, with an emphasis on the Laue lens. Finally, a sensitivity estimate is given, and the scientific perspectives it offers are discussed.

\section{Laue lens concept}
\begin{figure}[t]
\begin{center}
\includegraphics[width=0.38\textwidth]{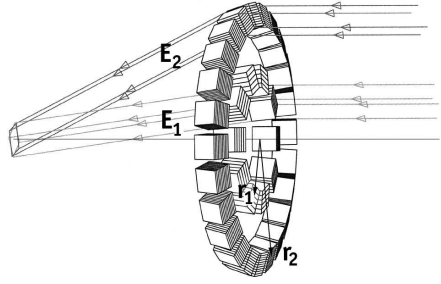}
\caption{Laue lens principle. Depending on the material and reflection of the crystals, E$_1$ and E$_2$ can be equal or not (see text for details). Crystals cross section measures 10$\times$10 mm$^2$ in the lens presented in this paper.}
\label{fig:LLprinciple}
\end{center}
\end{figure}

A Laue lens concentrates gamma-rays using Bragg diffraction in the volume of a large number of crystals arranged in concentric rings and accurately oriented in order to diffract radiation coming from infinity towards a common focal point (see e.g. Ref. \cite{lund.1992ft}). In the simplest design, each ring is composed of identical crystals with their axis of symmetry defining the optical axis of the lens (c.f. Figure \ref{fig:LLprinciple}).

Bragg's law links the angle $\theta_B$ between the ray's direction of incidence and the diffraction planes to the diffracted energy $E$ through the diffracting planes d-spacing $d_{hkl}$:
\begin{eqnarray}
2d_{hkl} \sin \theta_B = {hc \over E}  \: \: \Leftrightarrow \: \: 2d_{HKL} \sin \theta_B = n\, {hc \over E}
\end{eqnarray}
with $h$, $k$, $l$ being the Miller indeces defining the set of diffracting planes at work (another notation uses $H$, $K$, $L$ prime numbers and $n$ the order of diffraction), $h$ is the Planck constant and $c$ is the velocity of light in vacuum. Considering a focal distance $f$, the mean energy diffracted by a ring only depends on its radius $r$ and the d-spacing of its constituent crystals \cite{halloin.2005gd} :
\begin{eqnarray}
E = {hc \over 2 d_{hkl} \sin \left({1 \over 2} \arctan \left({r \over f} \right) \right)} \varpropto {f \over d_{hkl} \, r}~~~.
\label{eq:Efrd}
\end{eqnarray}

As we understand from equation (\ref{eq:Efrd}), if the variation of radius is compensated by a change in the d-spacing of the crystals, several rings can diffract the same energy adding up the effective area over a narrow energy band ($E_1=E_2$ in Fig. \ref{fig:LLprinciple}). However, if identical crystal and reflection (i.e. the same d-spacing) is employed on many adjacent rings, a continuous energy band is obtained (and $E_1 > E_2$ in Fig. \ref{fig:LLprinciple}). The Laue lens presented hereafter is a combination of these two principles.

More generally, the principle is applicable from $\sim$100 keV up to $\sim$1.5 MeV, limited at the low end by the fact that the crystals thickness maximizing the reflectivity becomes too small to allow their safe handling, and by the reflectivity that becomes too low at the other end. Although Laue lenses are better adapted to cover relatively narrow energy bandpasses, it is possible to get a sizable continuous effective area over several hundreds of keV, as demonstrated for instance in Ref. \cite{barriere.2009fk} and \cite{barriere.2007kx}.

\section{Proposed telescope design}
The proposed telescope consists of a Laue lens focusing in a $\sim$100-keV wide energy bandpass centered on 847 keV. At the focus is a compact Compton camera. A deployable mast maintains the two instruments 30 m apart, the Laue lens being attached to the spacecraft while the Compton camera is at the tip of the mast, far from the spacecraft mass. With the observatory placed at the second Sun-Earth Lagrangian point, this configuration would ensure a minimal instrumental background in the detector and cancel the differential gravitational forces that such a long structure would undergo in low Earth orbit, and this is basically the DUAL mission concept.  For the mission proposed to ESA, the compact Compton camera is left unshielded and thus sees nearly the whole sky all the time. In DUAL, the Compton camera acts simultaneously as focal plane for the lens and as all-sky monitor, surveying continuously the gamma-ray sky and accumulating more data on every source in the sky.

\begin{table}[t]
\begin{center}
\begin{minipage}[t]{.4\linewidth}

	\begin{small}
    \begin{tabular}{|l|c|}
\hline
Parameter & Value \\
\hline
\hline
Focal length (m) & 30.0 \\
Inner radius (cm) & 12.85 \\
Outer radius (cm) & 48.50 \\
Mass of crystals (kg) & 61.0 \\
Crystals size (mm$^2$) & 10 $\times$ 10 \\
Crystals thickness (mm) & 5.1 - 12.0 \\
Crystals mosaicity (arcsec) & 45 \\
Crystals inter-spacing (mm) & 0.5 \\
\hline
\end{tabular}
\end{small}

\end{minipage}
\hfill
\begin{minipage}[t]{.4\linewidth}

	\begin{small}
    \begin{tabular}{|l|c|}
\hline
Crystal material & Number \\
\hline
\hline
Rhodium  &  3193 \\
Silver &  1662 \\
Lead &   521\\
Copper &   348 \\
Germanium  &   76 \\
{\bf Total} & {\bf 5800} \\
\hline
\end{tabular}
\end{small}

\end{minipage}
\caption{Characteristics and composition of the Laue lens.}
\label{tab:lensparam}
\end{center}
\end{table}

\begin{figure}[t]
\begin{center}
\includegraphics[width=0.46\textwidth]{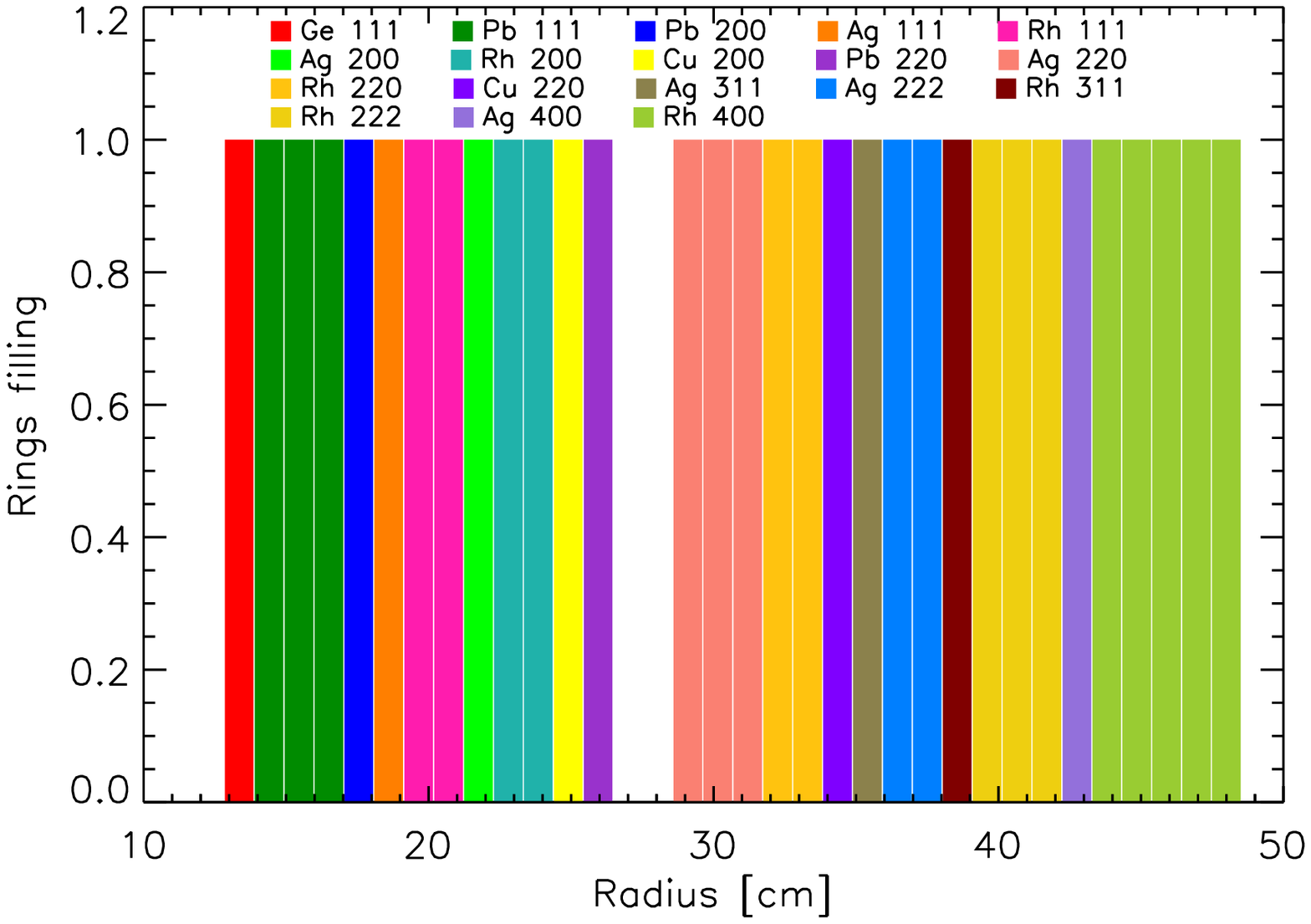}
\rotatebox{90}{
\includegraphics[width=0.33\textwidth]{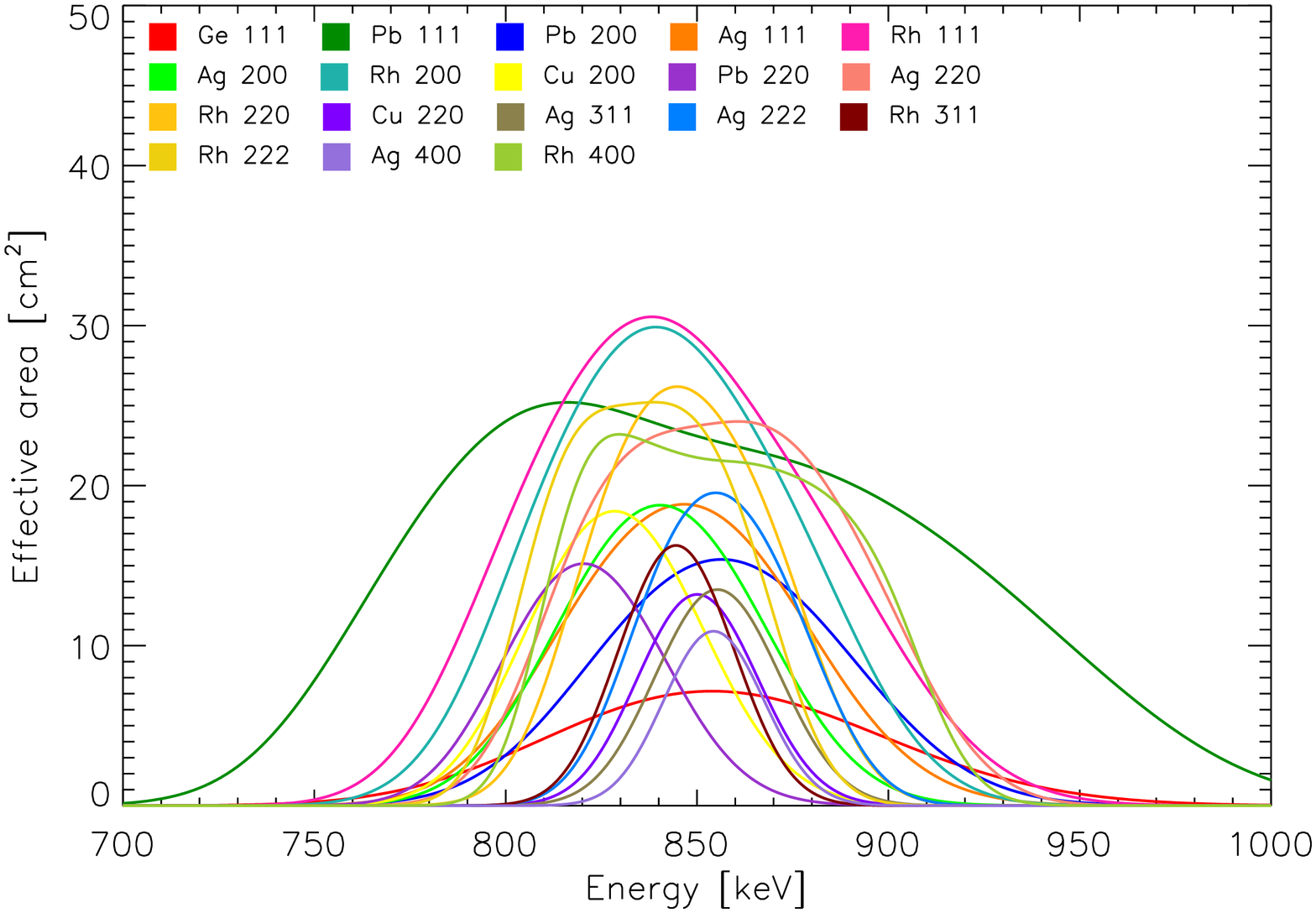}
}
\caption{\emph{Left:} Disposition of the crystals in the lens: composition (crystal material and reflection) and radius of each ring.
\emph{Right:} Close-up of the contribution of each group of crystal ring that are adding up to build the total effective area.}

\label{fig:crystdistrib}
\end{center}
\end{figure}

\subsection{Laue lens optic}

The Laue lens optic proposed for the DUAL mission is composed of 5800 crystal slabs of 10$\times$10 mm$^2$ arranged in 32 concentric rings, each populated by identical crystals. The crystals, Rh, Ag, Pb, Cu and Ge are mosaic crystals made of pure material. Table \ref{tab:lensparam} summarizes the parameters of the lens and details its composition, while Figure \ref{fig:crystdistrib} shows the arrangement of the crystals in rings.  Samples of every crystal material used in this design have been measured in a high energy photon beam (either using beamline ID15A at ESRF or instrument GAMS 5 at ILL, both in Grenoble, France), proving that they exist with the required mosaicity \cite{barriere_2010b}. Actually, there is no doubt that Ge and Cu are available for the realization of a Laue lens: 576 Ge crystals produced by IKZ (Berlin, Germany) were used for the CLAIRE Laue lens prototype \cite{von-ballmoos.2004sf}, and Cu crystals recently benefitted from an ESA-funded study, which established their reproducibility with constant quality in large quantities.  The ``new'' materials, Rh, Ag and Pb, are currently being studied at Space Sciences Laboratory in collaboration with Mateck GmbH (Juelich, Germany). The aim is to develop the method of producing a large series of crystals with constant quality, which encompasses growth of homogeneous ingots, orientation, quality check, cut, cut-induced damages removal (by acid etching) and final characterization. This activity is ongoing and should produce results around Fall 2011.

\begin{figure}[h!]
\begin{center}
\rotatebox{90}{
\includegraphics[width=0.33\textwidth]{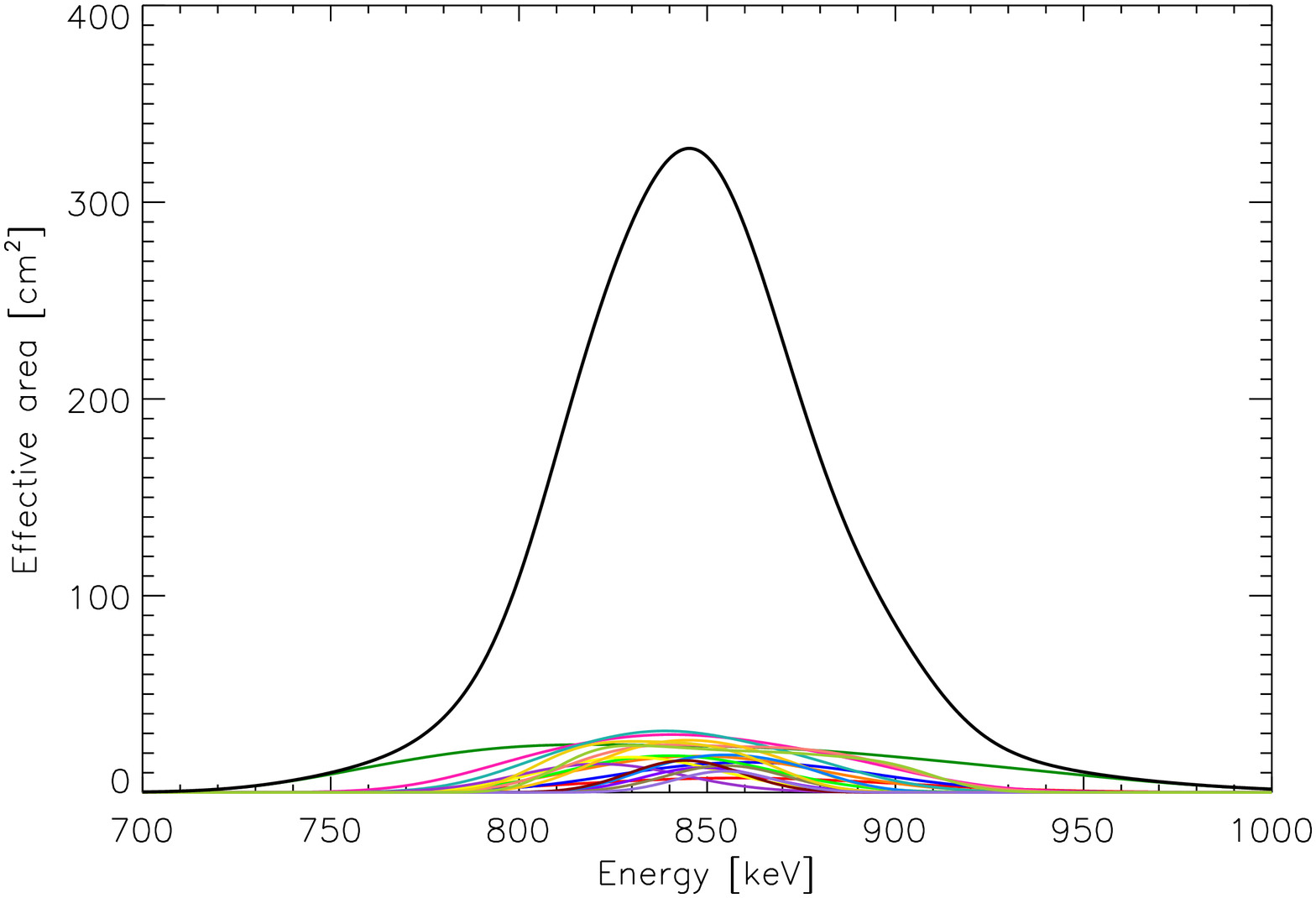}
}
\includegraphics[width=0.45\textwidth]{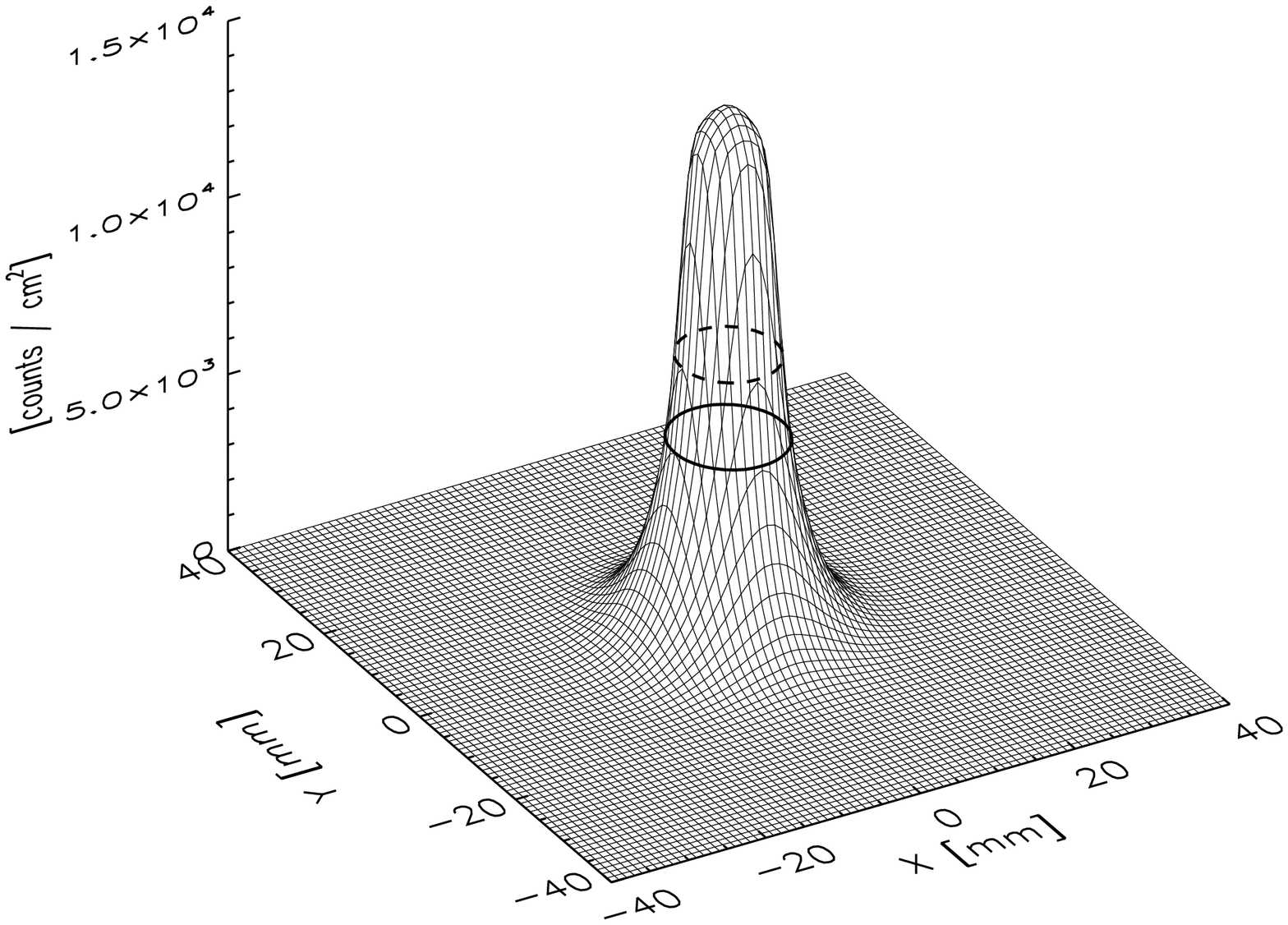}
\caption{\emph{Left:} Laue lens effective area for a point source on axis.
\emph{Right:} Focal spot for a point source on axis. The circular plain line shows the area optimizing the detection significance (assuming an homogeneous instrumental background), while the dashed line encompasses half the signal.}
\label{fig:LLEffArea}
\end{center}
\end{figure}

The calculation of the lens effective area (left panel of Figure \ref{fig:crystdistrib} and right panel of Figure \ref{fig:LLEffArea}) uses experimentally validated parameters to feed the mosaic crystal model (Darwin's model using the dynamical theory of diffraction \cite{Barriere:he5432}).  It takes into account 3 mm of equivalent aluminum for the substrate absorption and assumes crystals non-ideally oriented following a Gaussian distribution of $\sigma$=10 arcsec (c.f. next section). The peak effective area is 330 cm$^2$ at 847 keV and averages 318 cm$^2$ over a 26-keV wide band centered on 847 keV, which corresponds to the expected 3\% broadening of the line. As shown in the right panel of Figure \ref{fig:LLEffArea}, 60\% of the effective area is concentrated onto 1.45 cm$^2$, which represents a gain of 132.

\subsection{Crystal orientation precision requirement}

The aim of a Laue lens is to concentrate the maximum signal onto the smallest surface. The sensitivity of a Laue lens telescope is actually proportional to 
\begin{eqnarray}
f \propto {\sqrt{A_{focal spot}} \over A_{eff} \, \epsilon_{focal spot}}
\label{eq:sensitivity}
\end{eqnarray}
with $A_{focal spot}$ being the area on the detector that maximizes the detection efficiency, $\epsilon_{focal spot}$ is the fraction of signal that is encompassed in the area $A_{focal spot}$ and $A_{eff}$ the average effective area over the energy band of interest (a 0.03 $\times$ 847 keV wide band centered on 847 keV in the present case). We use this calculation to evaluate the tolerance on the orientation of the crystal about their most critical angle, that of the axis tangent to each ring, \emph{i.e.} the Bragg angle. Using this notation, Table \ref{tab:sens_des} shows the variation of the sensitivity with the misorientation of the crystals around their nominal angle. The distribution of misorientation is considered Gaussian. A misorientation of $\sigma_{misorientation}$ = 30 arcsec results in a factor of 2 in sensitivity loss. Thus, the goal for the assembly of the lens is 10 arcsec, and the requirement is 20 arcsec.

\begin{table}[t]
\begin{small}
\begin{center}
\begin{tabular}{|x{2.5cm}|x{2.5cm}|x{2.5cm}|x{2.5cm}|p{2.5cm}|}
\hline
$\sigma_{misorientation}$ (arcsec) & $A_{focal spot}$ (cm) & $\epsilon_{focal spot}$ & A$_{eff}$ (cm$^2$)& Sensitivity (Normalized) \\
\hline
\hline
0 & 1.41 & 0.628 & 329.9 & 1 \\
10 & 1.45 & 0.602 & 318.9 & 1.10 \\
20 & 1.67 & 0.567 & 282.5 & 1.41 \\
30 & 2.1 & 0.525 & 244.5 &  1.98 \\
\hline
\end{tabular}

\end{center}
\end{small}
\caption{Variation of the sensitivity of the telescope with the misorientation of the crystals around their nominal angle \label{tab:sens_des}
}
\end{table}

\subsection{Field of view and pointing precision requirement}
\begin{figure}[t]
\begin{center}
\includegraphics[width=0.4\textwidth]{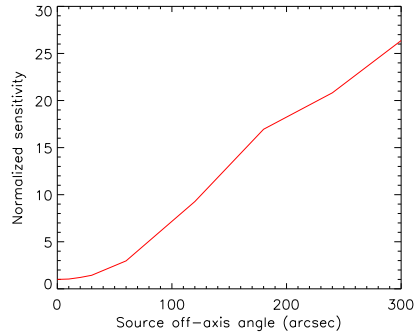}
\caption{Relative sensitivity evolution with the point source angular distance to the lens optical axis.}
\label{fig:LLFoV}
\end{center}
\end{figure}

Calculated according to the same principle as in the previous section, Figure \ref{fig:LLFoV} shows the variation of the sensitivity (normalized to the on-axis value) for a point source, as a function of its angular distance to the optical axis of the lens (\emph{i.e.} the off-axis angle). As one can see, the sensitivity drops extremely quickly with the off-axis angle. An off-axis angle of 125 arcsec induces a sensitivity loss of an order of magnitude. Indeed this lens is not designed to make an image or to survey the sky. It is designed to observe point sources of known position, as the peak of the 847 keV line light curve arises $\sim$ 50 days after the WD explosion, well after the optical light curve has reached its maximum intensity.

From this curve, one can determine the pointing accuracy required. A decrease of sensitivity by a factor of two occurs when the source is 42 arcsec off-axis.  Thus, the requirement is to keep the source no more than 30 arcsec from the on-axis position (sensitivity loss factor of 1.4), with a goal of 15 arcsec (sensitivity loss factor of 1.2).

\subsection{Substrate and thermal control}
For such a small lens (external diameter lower than 1 m), a monolithic substrate is preferred as it would save the weight of module-to-module attachments and prevent any loss of area at the interfaces, insuring a maximum packing factor for the crystals. An ideal material to build the substrate is silicon carbide, as it is extremely lightweight, allows for precise machining, and is a good thermal conductor. Based on existing devices, one can estimate the weight of the substrate to be lower than 10 kg for the proposed Laue lens \cite{devilliers.2006fk}.

The thermal stability requirement derives from the crystals that should keep their orientation within $\pm$10 arcsec with respect to the lens optical axis. It means that any warping should not induce any slope of more than 10 arcsec with respect to the nominal plane. During the MAX \cite{barriere.2005fp} pre-phase A study at CNES (the French Space Agency), the thermal control of the Laue lens was investigated \cite{hinglais.2005sf}. It showed that a cocoon of multilayer insulator surrounding the lens combined with a few heaters and thermistors on the lens structure are sufficient to limit the thermal gradient to 2$^\circ$C across the lens for any sun exposition, which in the case of MAX would prevent any warping worst than 10 arcsec. The proposed Laue lens has a much simpler and smaller structure that MAX's, so it is reasonable to believe that this result can be safely applied to the present case. Hence, a simple thermal control mainly based on passive insulation should allow the lens to maintain its assembly temperature within 2$^\circ$C, which should be enough to ensure nominal performance.

\subsection{Leads for performance improvement}

\begin{figure}[t]
\begin{center}
\includegraphics[width=0.4\textwidth]{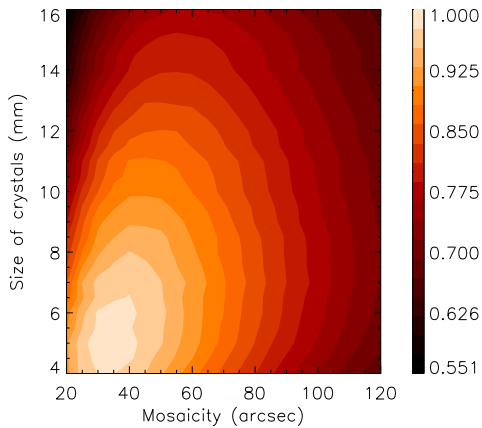}
\caption{Relative sensitivity evolution with crystals size (assumed square shaped) and mosaicity. This calculation takes into account crystals non ideally oriented following a Gaussian distribution of $\sigma$=10 arcsec.}
\label{fig:mos_size}
\end{center}
\end{figure}

The replacement of the silver crystals by gold would result in an increase of the effective area of 5\% and a decrease of the crystal mass of 6\%. Additionally, two rings populated with InSb crystals could be added on the inside of the lens, adjacent to the Ge ring. This would add $\sim$1kg to the total mass and would increase the effective area by $\sim$ 3.5\% (\emph{i.e.} 12 cm$^2$). Additionally, smaller crystals with slightly less mosaicity would increase the sensitivity as it would decrease the size of the focal spot on the detector. Figure \ref{fig:mos_size} shows that crystals of 8$\times$8 mm$^2$ with a mosaicity of 40 arcsec (instead of 10 $\times$10 mm$^2$ and 45 arcsec) would increase the sensitivity by $\sim$ 5\%.

It appears that although there are possibilities for sensitivity improvement with the present design, they are minor. If a major improvement is to be achieved, the only way, to our knowledge, is to increase the focal length. A longer focal length would allow more area for the crystals, leading to a sizable increase of the effective area at the cost of an increase of the lens mass. As an example, a similar Laue lens having a focal length of 40 m has been designed. It results in an increase of sensitivity of 33\%, for a crystal mass of 67 kg (7000 crystal slabs of 10$\times$10 mm$^2$) and an outer radius of 112 cm.

\subsection{Focal plane}
The focal plane considered for the performance estimates (c.f. next section) is the All-Sky Compton Imager featured in the DUAL mission. This camera is made of 45 planar cross-stripped high purity Ge detectors, similar to that developed for the Nuclear Compton Telescope \cite{bellm.2010kx}, arranged in 5 layers of 3$\times$3 elements. Each detector measures 10$\times$10$\times$1.5 cm$^3$ with 2-mm pitch strips, allowing event positioning within a volume of 1.6 mm$^3$. These detectors combine spatial resolution and excellent spectral resolution, making them perfect elements for a Compton camera. 

It is to be noted however that there is no need for such a big focal plane for the Laue lens alone. This design was driven by the need to have a powerful all-sky imager. Regarding the Laue lens, the same performance would be obtained by a single stack of 5 HPGe elements surrounded by 4 additional elements acting as side walls.

\section{Performance - conclusions}
The sensitivity of the instrument presented in this paper has been calculated using the Megalib code \cite{zoglauer.2006uq} in order to account for Compton reconstruction of the events in the focal plane, thus applying ``electronic collimation'' by rejecting events whose cone of possible incidence direction do not intersect the lens and whose first interaction lies in the PSF of the lens. The resulting 3-$\sigma$ sensitivity is 1.8$\times$10$^{-6}$ ph s$^{-1}$ cm$^{-2}$ for 10$^6$ s observation time of a 3\% broadened line. This is about 30 times more sensitive than INTEGRAL/SPI. With such tremendous sensitivity, SNe Ia can be detected out to $\sim$40 Mpc, which means dozens of detection per year, with \emph{at least} one with high significance. Enough to make a real breakthrough in our understanding of these events.

The Laue lens concept presented in this paper weighs about 80 kg (including margin), and is based on existing crystals and existing substrate technology. Different mounting methods for the crystals are being developed in Toulouse (France), Ferrara (Italy) and more recently in Berkeley (USA), driving important progresses in the technology readiness level. According to industrial partners, the deployable mast technology is not a critical point for this kind of mission, as the NuSTAR mission \cite{harrison.2010fk} with soon demonstrate with its 10-m long mast. All the ingredients are getting ready to finally be able to construct a telescope powerful enough to address the true nature of SNe Ia.

 \acknowledgments
The authors thank Dr. Andreas Zoglauer for his contribution in determining the sensitivity of the proposed telescope through an end to end calculation including the Compton reconstruction of the events in the focal plane, and the simulation of the instrumental background for an L2 orbit assuming the DUAL all-sky Compton telescope as focal plane.


\end{document}